\documentclass[twocolumn]{svjour3}          
\smartqed  
\usepackage{graphicx}
\journalname{Tribology Letters}
\newcommand{\sgn}{\mathrm{sgn}}
\newcommand{\dd}{\mathrm{d}}
\newcommand{\rev}{\textcolor{black}}
\usepackage{color,amssymb}

\begin{document}

\title{Relaxation tribometry: a generic method to identify the nature of contact forces
}


\author{Alain Le Bot       \and
            Julien Scheibert \and
            Artem A. Vasko      \and
            Oleg M. Braun 
}


\institute{
A. Le Bot \and J. Scheibert 
\at Univ Lyon, Ecole Centrale de Lyon, ENISE, ENTPE, CNRS, Laboratoire de Tribologie et Dynamique des Syst\`emes LTDS UMR5513, F-69134, Ecully, France \\\email{alain.le-bot@ec-lyon.fr}
\and
A.A. Vasko \and O.M. Braun
\at Institute of Physics, National Academy of Sciences of Ukraine, 46 prospect Nauki, Kiev, 03028, Ukraine
}

\date{Received: date / Accepted: date}

\maketitle

\begin{abstract}
Recent years have witnessed the development of so-called relaxation tribometers, the free oscillation of which is altered by the presence of frictional stresses within the contact. So far, analysis of such oscillations has been restricted to the shape of their decaying envelope, to identify in particular solid or viscous friction components. Here, we present a more general expression of the forces possibly acting within the contact, and retain six possible, physically relevant terms. Two of them, which had never been proposed in the context of relaxation tribometry, only affect the oscillation frequency, not the amplitude of the signal. We demonstrate that each of those six terms has a unique signature in the time-evolution of the oscillation, which allows efficient identification of their respective weights in any experimental signal. We illustrate our methodology on a PDMS sphere/glass plate torsional contact.

\keywords{Relaxation tribometer \and Damped oscillations \and Amplitude decay curve \and Frequency shift \and Nonlinear contact forces \and Two-times averaging method}
\end{abstract}

\section{Introduction}
\label{sec:intro}

The energy dissipated during relative motion of solid surfaces in contact corresponds to the work of the friction force. It is therefore appealing to measure friction forces without force sensors, just from the energy decay that they induce in a frictional system. This is precisely what the so-called relaxation tribometry is about. The basic idea is to place a tribological interface in an oscillator, provide the latter with a certain amount of initial mechanical energy, and let it oscillate and relax back to its equilibrium position. The time-rate of such a relaxation informs about the amplitude of the friction force, while the envelope of the decaying oscillation characterizes the type of dissipative mechanism involved.

The idea of measuring a viscous damping coefficient by monitoring the decay of vibration of an oscillator dates back to Rayleigh, who described the logarithmic decrement technique in his famous treatise~\cite[p.46]{bib:rayleigh}. But relaxation tribometry actually starts with the remark that a velocity-independent friction coefficient can also be measured from the time-decay of the envelope of the vibration~\cite{bib:lorenz1924,bib:kimball1929}. When both friction and viscous dissipation are present, the solution of the governing equation of an oscillator and its amplitude decay curve was found by Markho~\cite{bib:markho1980}. This led Feeny and al.~\cite{bib:feeny1996,bib:feeny1998} to extend the decrement method to measure simultaneously viscous and friction coefficients,  while Wu and al.~\cite{bib:wu2007} proposed to apply the method to nonlinear viscous damping. Rigaud \textit{et al.}~\cite{bib:rigaud2010} performed similar simultaneous measurements, not from the amplitude but from the energy decay, on contacts lubricated by water-glycerol solutions.

Recently, renewed interest for relaxation tribometry has emerged as a unique tool to measure low forces efficiently and accurately~\cite{bib:majdoub2013,bib:majdoub2014,bib:belin2018}. The reason is that the smaller the friction force, the smaller the decrement, the more measurable oscillations before rest and thus the more data available to estimate the force.

Moreover, the method is so accurate that non-conven\-tional behaviours, neither purely frictional nor purely viscous, could be detected. On the one hand, nonlinear dissipative forces have been identified and measured~\cite{bib:majdoub2015}. On the other hand, a progressive shift of the oscillation frequency has been observed with a rotational tribometer~\cite{bib:vasko2018}. Both observations confirm that tribological interfaces are complex~\cite{bib:vakis2018} and thus cannot be fully described with a simple combination of viscous and friction coefficients. A sliding contact, in translation or torsion, is in particular made of a multitude of micro-contacts~\cite{bib:bavouzet2010,bib:sahli2018}, possibly implying different materials~\cite{bib:smerdova2012}. It can also be divided into coexisting slip and stick zones~\cite{bib:chateauminois2010,bib:prevost2013}.
The question thus arises of which nonlinear forces can actually be detected by relaxation tribometry.

In this paper, we propose a generic description of the type of signals recovered using relaxation tribometry. By applying systematically the two-times averaging method already used in~\cite{bib:rigaud2010}, we describe the characteristic signatures of six relevant contact forces, in terms not only of the decay of their amplitude, but also of the frequency shift.


\section{Theoretical development}
\label{sec:theory}

\begin{figure}[ht!]
\includegraphics[scale=1.0]{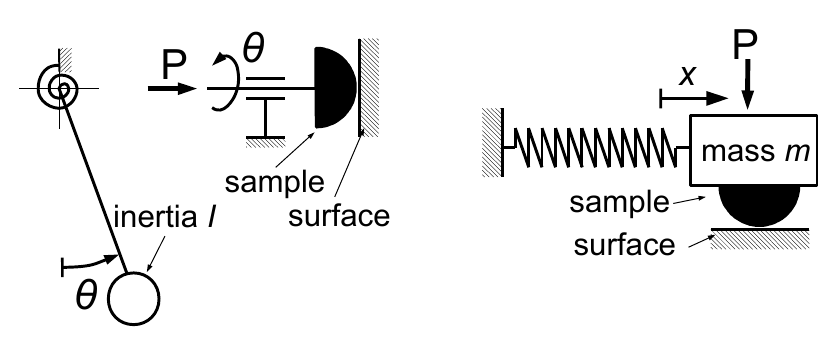}
\caption{Left: principle of a torsional pendulum with a sphere/plane contact at the extremity of the rotation axis. Right: principle of a translational oscillator with a sphere in contact with a plane and attached to the moving mass.}
\label{fig:pendulum}
\end{figure}

The principle of a relaxation tribometer consists in observing the free vibration of an oscillator equipped with a sample rubbing on a surface. The decay of vibration, due to friction and other sources of dissipation, holds information on contact forces acting on the sample. Thus, a simple measurement of the motion allows to extract the friction force without using a force probe. \rev{Note that relaxation tribometry, in essence, probes the transient (as opposed to steady sliding) response of a frictional interface over the characteristic time scale of the oscillator's period.} Two types of tribometer have been reported in the literature~\cite{bib:rigaud2010,bib:vasko2018} depending on the kinematics followed by the rubbing sample:  translation or torsion (see Fig.~\ref{fig:pendulum}).
 
In this section, we will argue on the basis of a torsional relaxation tribometer (Fig.~\ref{fig:pendulum}, left), with no loss of generality. For a translational relaxation tribometer (Fig.~\ref{fig:pendulum}, right), one should simply replace the angle $\theta$ with the dimensionless position $m\omega_0^2x/P$, and the angular speed $\dot{\theta}$ with the dimensionless  speed $m\omega_0^2\dot{x}/P$ where $m$ is the moving mass, $\omega_0$ the natural frequency and $P$ the normal load.

In a torsional relaxation tribometer, an axisymmetric sample is pressed against a surface by a load $P$ and is submitted to torsional oscillations. The governing equation is:
\begin{equation}\label{eqn:mvt}
\ddot \theta + \omega_0^2 \theta = -\omega_0^2 f(\theta,\dot \theta)
\end{equation}
where $\omega_0$ is the contactless natural frequency of the oscillator and $f(\theta,\dot\theta)$ a dimensionless force induced by the sample rubbing on the surface and/or by a non-ideal behaviour of the pendulum. Here $f=-M/(I\omega_0^2)$ where $M$ is the external torque on and $I$ the total inertia of the pendulum. If we had considered a translational tribometer, then $f=-T/P$ where $T$ is the transverse force.  The origin $\theta=0$ is conventionally fixed at equilibrium. By construction, the pendulum is symmetric. The force $f$ is odd with respect to  parity and time reversal, so that:
\begin{equation}\label{eqn:symmetry}
f(-\theta,-\dot\theta) = -f(\theta,\dot\theta).
\end{equation}
The reactive force $f$ may still take a wide variety of forms, two special cases being of particular interest.

On the one hand, when the force is a dissipative reaction of the contact (typically friction), it is usually assumed to depend on the sliding speed $\dot\theta$ only, and not on the position $\theta$ and Eq.~(\ref{eqn:symmetry}) imposes $f(-\dot\theta) = - f(\dot\theta)$. For instance, a linear viscous force $f(\dot\theta) \propto \dot\theta$ as well as any drag force $f(\dot\theta) \propto \dot\theta^n$ with $n$ odd, satisfies this condition. Another example is the solid friction force $f(\dot\theta) \propto \sgn(\dot\theta)$, which is discontinuous at zero. The general form including these examples and matching the dissipative condition $-f(\dot\theta) \dot \theta < 0$ is an odd pseudo-polynomial of the type $f(\dot\theta) = \lambda \sgn(\dot\theta) + 2\zeta\dot\theta/\omega_0 + \delta \dot\theta^2\sgn(\dot\theta)/\omega_0^2  + o(\dot\theta^2)$, where $\lambda$, $\zeta$ and $\delta$ are positive dimensionless constants.


On the other hand, when $f$ is conservative, it derives from a potential and consequently does not depend on $\dot\theta$. Since $f$ is an odd function of $\theta$ (by Eq.~\ref{eqn:symmetry}) and continuous at zero, a pseudo-polynomial series expansion gives $f(\theta) = \alpha \theta + \nu \theta^2 \sgn(\theta) + \epsilon \theta^3 + o(\theta^3)$ where the constants may be positive or negative. 

Limiting ourselves to the above orders (2 in $\dot{\theta}$ and 3 in $\theta$), we are therefore left with six terms to study: a constant friction force $f=\lambda\sgn(\dot\theta)$, a linear viscous force $f=2\zeta \dot\theta / \omega_0$, a quadratic dissipative force $f=\delta\dot\theta^2\sgn(\dot\theta)/\omega_0^2$, a linear elastic force $f=\alpha \theta$, a quadratic elastic force $f=\nu\theta^2\sgn(\theta)$, and a cubic elastic force $f=\epsilon\theta^3$.

For weakly nonlinear systems -- when $f$ is small -- the solution of Eq.~(\ref{eqn:mvt}) may be approximated by:
\begin{equation}\label{eqn:sol}
\theta(t) = a(t) \cos \left[ \omega_0 t + \phi(t) \right]
\end{equation}
where $a(t)$ and $\phi(t)$ are slowly varying functions, {\textit{i.e.} over times much larger that $2\pi/\omega_0$. The magnitude $a(t)$ and phase $\phi(t)$ will now be obtained by applying the two-times averaging method~\cite{bib:strogatz,bib:nayfeh,bib:belhaq2013} (see Appendix for a brief review of the method).

In practice, the force $f$ acting on the pendulum is not known. One has to determine it by observing the behaviour of the pendulum. Thus, the time evolution of the magnitude, $a(t)$, and of the phase, $\phi(t)$, will constitute the only available information to identify the relevant terms in the right-hand side of Eq.~(\ref{eqn:mvt}).

\subsection{Decaying envelope}

The magnitude $a(t)$ constitutes the envelope of vibration, the decay of which indicates the energy lost in the sample. When the six forces are present simultaneously and are small, the two-times averaging method applies. The time derivative of the magnitude, $\dot a$, is given in Appendix, Eq.~(\ref{eqn:hsin}) where the right-hand side is obtained  by summing all terms $\langle h\sin\varphi\rangle$ given in Table~\ref{tab:1}. The result is:
\begin{equation}\label{eqn:equa_diff_a}
\dot a = - \frac{2}{\pi} \omega_0 \lambda - a \omega_0 \zeta  - \frac{4a^2}{3\pi} \omega_0 \delta.
\end{equation}
This is an ordinary differential equation of first order on $a$. Let us remark that the constants $\alpha$, $\nu$, and $\epsilon$ do not appear.

As observed in \cite{bib:lorenz1924}, a pure constant friction force $f=\lambda\sgn(\dot\theta)$ imposes a linear decreasing of the successive local maxima. The equation of the envelope is obtained by integrating (\ref{eqn:equa_diff_a}) with $\zeta=\delta=0$:
\begin{equation}\label{eqn:env_friction}
a(t) = a(0)-\frac{2 \lambda}{\pi}  \omega_0 t.
\end{equation}
An interesting consequence of Eq.~(\ref{eqn:env_friction}) is that since $a(t)\geq 0$, the vibration always stops after a finite duration equal to $a(0) \pi / 2\lambda \omega_0$. This is illustrated in Fig.~\ref{fig:decay}, top for initial angle $\theta(0)=1$ and speed $\dot\theta(0)=0$.

\begin{figure}
\includegraphics[scale=0.45]{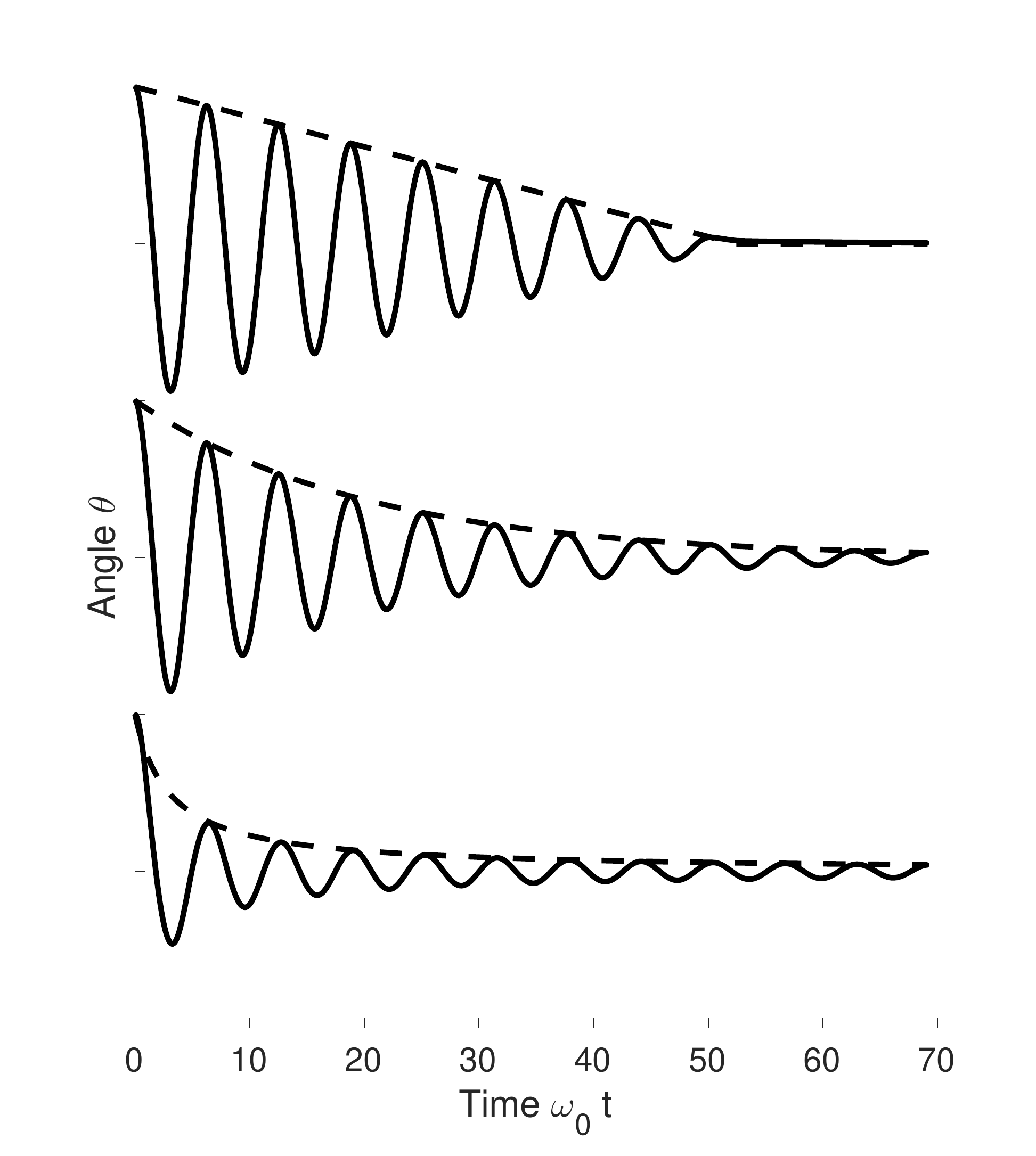}
\caption{Time evolution of angle $\theta$ for various forces $f$. Top: case of constant friction force $\lambda\sgn(\dot\theta)$ for $\lambda=0.03$. Middle: case of linear viscous force $2\zeta \dot\theta/\omega_0$ for  $\zeta=0.05$. Bottom: case of quadratic dissipative force $\delta\dot\theta^2\sgn(\dot\theta)/\omega_0^2$ for $\delta=0.9$. Solid line: numerical solution to Eq.~(\ref{eqn:mvt}) with initial conditions $\theta(0)=1$, $\dot\theta(0)=0$. Broken line: envelope by Eqs.~(\ref{eqn:env_friction}), (\ref{eqn:env_viscous}), and (\ref{eqn:env_quadDiss}).}
\label{fig:decay}
\end{figure}

A pure linear viscous force $f=2\zeta \dot\theta/\omega_0$ imposes an exponentially decreasing magnitude \cite{bib:rayleigh}. The equation of the envelope is from (\ref{eqn:equa_diff_a}) with $\lambda=\delta=0$:
\begin{equation}\label{eqn:env_viscous}
a(t) = a(0) \exp(-\zeta\omega_0 t).
\end{equation}
Thus the pendulum vibrates forever (Fig.~\ref{fig:decay}, middle).

The case of a pure quadratic dissipative force of type $f=\delta\dot\theta^2\sgn(\dot\theta)/\omega_0^2$ is more original, and to our knowledge has only been investigated in \cite{bib:majdoub2015}. The integration of (\ref{eqn:equa_diff_a}) with $\lambda=\zeta=0$ gives:
\begin{equation}\label{eqn:env_quadDiss}
a(t) = \frac{1}{\left[\frac{1}{a(0)}+\frac{4\delta}{3\pi} \omega_0 t \right]}
\end{equation}
This result is illustrated in Fig.~\ref{fig:decay}, bottom.

The three other terms, linear elastic force $f=\alpha \theta$, quadratic elastic force $f=\nu\theta^2\sgn(\theta)$ and cubic elastic force $f= \epsilon\theta^3$, are conservative. Consequently, the corresponding magnitude $a(t)=a(0)$ is constant over large time scales. This is consistent with the fact that $\alpha$, $\nu$, and $\epsilon$ do not appear in Eq.~(\ref{eqn:equa_diff_a}).

\subsection{Varying frequency}
The time evolution of the phase $\phi(t)$ is a typical nonlinear effect. The instantaneous frequency is given by:
\begin{equation}
\omega = \frac{\dd}{\dd t} \left[ \omega_0 t + \phi(t) \right] = \omega_0 + \dot\phi(t)
\end{equation}
where $\dot\phi$ is the time-derivative of $\phi$.

There exists analytical results for the value of the angular frequency in the presence of some of the forces considered here. The case of a linear elastic force $f=\alpha \theta$ is rather trivial since one can write Eq.~(\ref{eqn:mvt}) as $\ddot \theta + \omega_0^2 (1+\alpha) \theta =0$, so that the angular frequency of the oscillation has a constant value $\omega_0 \sqrt{1+\alpha}$. If a viscous damping force $f=2\zeta \dot\theta/\omega_0$ is added, the frequency of the oscillation remains constant during the oscillation, but with the value $\omega_{\infty}=\omega_{0}\sqrt{1+\alpha}\sqrt{1-\zeta^2}$. It has been shown in~\cite{bib:feeny1996} that this conclusion remains true if a constant friction force $f=\lambda\sgn(\dot\theta)$ is added.

For the other three forces, we solved the evolution of the angular frequency using the two-times averaging method. Equation~(\ref{eqn:hcos}) of Appendix gives $a\dot\phi$ as a linear combination of all terms $\langle h\cos\varphi\rangle$ given in Table~\ref{tab:1}. The frequency $\omega=\omega_0 + \dot\phi$ is then:
\begin{equation}\label{eqn:equa_omega}
\omega = \omega_0 \left( 1 + \frac{\alpha}{2} + \frac{4a}{3\pi} \nu  + \frac{3 a^2}{8}\epsilon  \right).
\end{equation}

Note that, according to the two-times averaging method, Eqs.~(\ref{eqn:equa_diff_a}) and~(\ref{eqn:equa_omega}) correspond to a linear approximation in all small terms $\lambda$, $\zeta$, $\delta$, $\alpha$, $\nu$ and $\epsilon$. In particular, to this degree of approximation, $\sqrt{1-\zeta^2}\simeq 1$ and $\sqrt{1+\alpha}\simeq 1+ \alpha / 2$, so that $\omega_{\infty}\simeq\omega_{0}(1+ \alpha / 2)$.

Also note that the constants $\lambda$, $\zeta$, and $\delta$ do not appear in Eq.~(\ref{eqn:equa_omega}). In particular, although highly nonlinear, the dissipative quadratic force (amplitude $\delta$) does not induce any time variation of the frequency (Fig.~\ref{fig:periodQuadraticDissipative}).

\begin{figure}[tb!]
\includegraphics[scale=0.45]{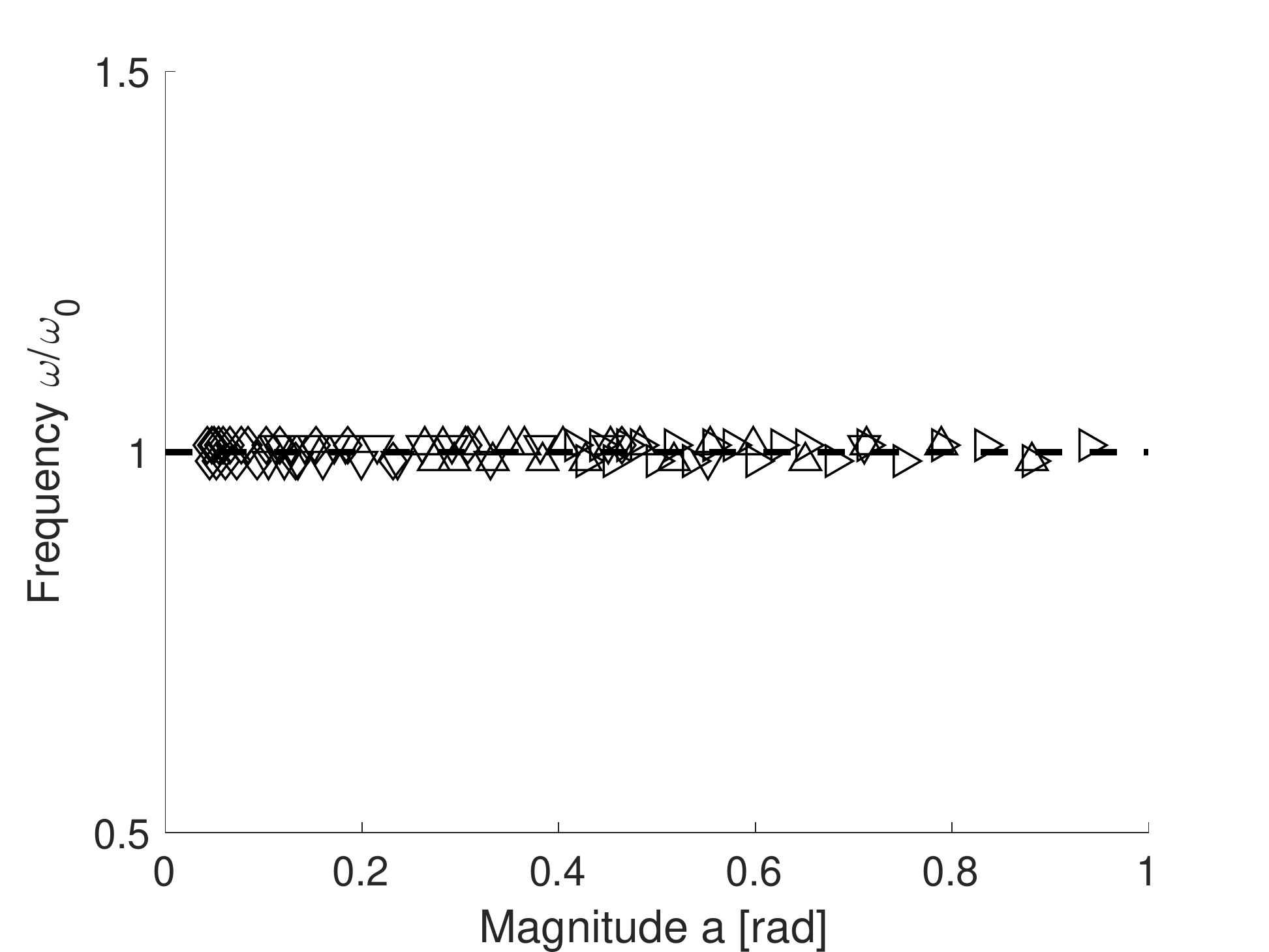}
\caption{Evolution of frequency with amplitude for a pure quadratic dissipative force $\delta\dot\theta^2\sgn(\dot\theta)/\omega_0^2$. Symbols: numerical solution to Eq.~(\ref{eqn:mvt}) with initial conditions $\theta(0)=1$, $\dot\theta(0)=0$ (one value per half-period) with $\delta$=0.05 ($\vartriangleright$), 0.1 ($\bigtriangleup$), 0.3 ($\bigtriangledown$) or 0.8 ($\lozenge$). Broken line: Eq.~(\ref{eqn:equa_omega}).}
\label{fig:periodQuadraticDissipative}
\end{figure}

The situation is different for elastic forces. In the case of a pure quadratic elastic force $f=\nu\theta^2\sgn(\theta)$, Eq.~(\ref{eqn:equa_omega}) gives $\Delta\omega = \omega - \omega_\infty = 4\nu a \omega_0 / (3\pi)$.
Thus, a quadratic elastic force exhibits a linear variation of frequency versus magnitude of envelope. The sign of $\nu$ controls the type of frequency evolution: an experiment exhibiting an increasing frequency with increasing time, \textit{i.e.} with decreasing amplitude of the oscillation corresponds to $\nu<0$. Conversely, if the period of the pendulum increases with time, then $\nu>0$. Those results are demonstrated on Fig.~\ref{fig:periodQuadraticConservative}, which successfully compares Eq.~(\ref{eqn:equa_omega}) with simulation results.

\begin{figure}[tb!]
\includegraphics[scale=0.45]{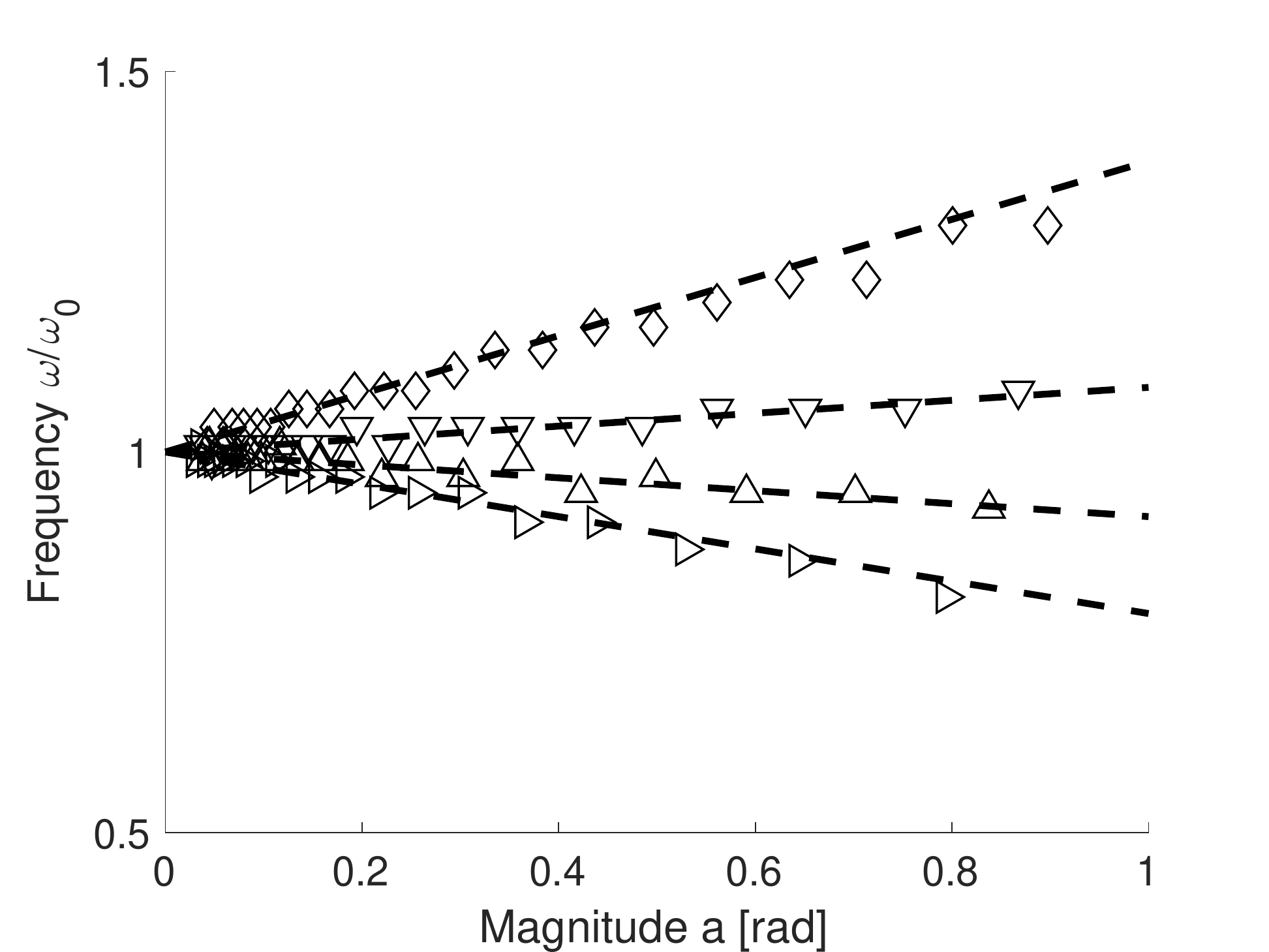}
\caption{Evolution of frequency with amplitude for a pure quadratic elastic force $\nu\theta^2\sgn(\theta)$. Symbols: numerical solution to Eq.~(\ref{eqn:mvt}) with initial conditions $\theta(0)=1$, $\dot\theta(0)=0$ (one value per half-period) with $\zeta=0.05$ and $\nu$=-0.5 ($\vartriangleright$), -0.2 ($\bigtriangleup$), 0.2 ($\bigtriangledown$), 0.9 ($\lozenge$). Broken lines: Eq.~(\ref{eqn:equa_omega}).}
\label{fig:periodQuadraticConservative}
\end{figure}

\begin{figure}[tb!]
\includegraphics[scale=0.45]{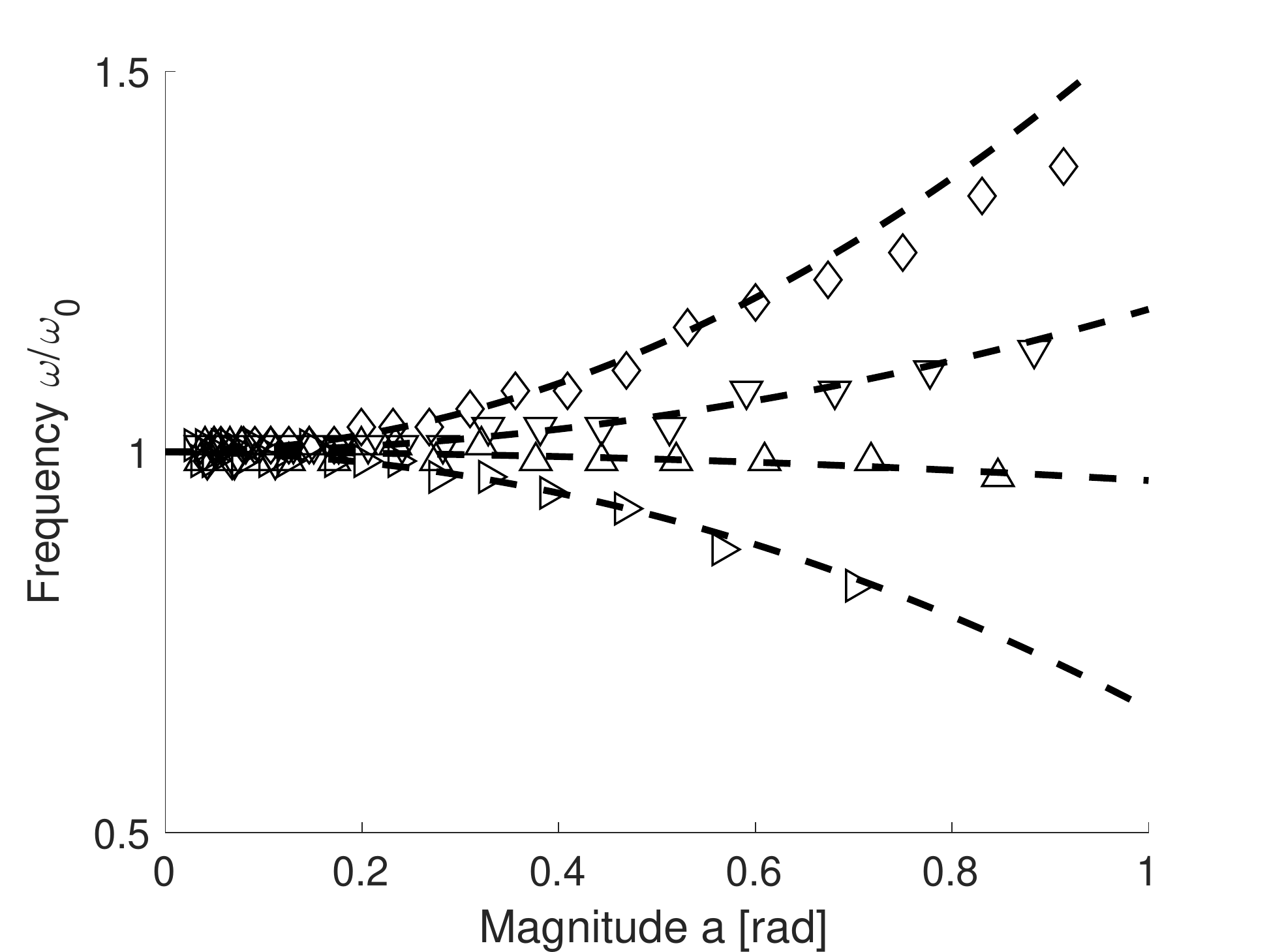}
\caption{Evolution of frequency with amplitude for a pure cubic elastic force $\epsilon\theta^3)$. Symbols: numerical solution to Eq.~(\ref{eqn:mvt}) with initial conditions $\theta(0)=1$, $\dot\theta(0)=0$ (one value per half-period) with $\zeta=0.05$ and $\epsilon$=-0.9 ($\vartriangleright$), -0.1 ($\bigtriangleup$), 0.5 ($\bigtriangledown$), 1.5 ($\lozenge$). Broken lines: Eq.~(\ref{eqn:equa_omega}).}
\label{fig:periodCubicConservative}
\end{figure}

Finally, the case of a pure cubic elastic force, which corresponds to the well known Duffing oscillator, is also solved by Eq.~(\ref{eqn:equa_omega})~\cite{bib:strogatz}. The frequency shift is now given by $\Delta\omega = \omega - \omega_\infty = 3\epsilon a^2 / 8$, where the variation of frequency is quadratic in $a$. As for the quadratic elastic force, the sign of $\epsilon$ may be determined by observing whether the pendulum experiences an increasing ($\epsilon>0$) or decreasing ($\epsilon<0$) period as time increases. Again, those results are found in good agreement with simulation results (Fig.~\ref{fig:periodCubicConservative}) although small discrepancies are visible when $\epsilon$ is not small enough.

\subsection{Identification of force strength}\label{sec:strength}
To get effective values of the six coefficients $\lambda$, $\zeta$, $\delta$, $\alpha$, $\nu$, and $\epsilon$, we start from the two curves $a(t)$ and $\omega(t)$, both assumed to be known with a sufficient precision. The method to extract $a(t)$ and $\omega(t)$ from the full measurement of $\theta(t)$ is by no way important at this stage. We admit that the magnitude $a$, its time-derivative $\dot a$, and the frequency $\omega$ are known at a finite number of times (see Fig.~\ref{fig:principleMeasure}). Let $a_i = a(t_i)$, $\dot a_i = \dot a(t_i)$ and $\omega_i=\omega(t_i)$ be these values, for $i=1,\dots,n$. 

\begin{figure}[htb!]
\includegraphics[scale=1.0]{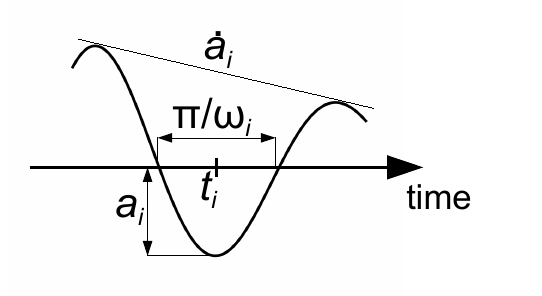}
\caption{Estimation of magnitude $a_i$, slope of magnitude $\dot a_i$, and frequency $\omega_i$ at time $t_i$.}
\label{fig:principleMeasure}
\end{figure}


Let us first determine $\lambda$, $\zeta$, and $\delta$. By Eq.~(\ref{eqn:equa_diff_a}):
\begin{equation}
\dot a_i = - \frac{2}{\pi} \omega_0\lambda - a_i \omega_0\zeta - \frac{4a_i^2}{3\pi} \omega_0\delta \quad\quad i=1,2,\dots.
\end{equation}
Assembling these equations in a matrix form, we get:
\begin{equation}\label{eqn:epsilon_system}
\omega_0\left(\begin{array}{ccc} -\frac{2}{\pi} & -a_1 & -\frac{4a_1^2}{3\pi} \\  \dots &\dots & \dots \\ -\frac{2}{\pi} & -a_n & -\frac{4a_n^2}{3\pi}  \end{array} \right)  \left(\begin{array}{c} \lambda \\ \zeta \\ \delta \end{array} \right) = \left(\begin{array}{c} \dot a_1 \\ \dots \\ \dot a_n \end{array} \right).
\end{equation}
This equation constitutes an overdetermined system of linear equations. The solution in the least mean square sense is given by:
\begin{equation}\label{eqn:epsilon_dissipative}
\left(\begin{array}{c} \lambda \\ \zeta \\ \delta \end{array} \right) = \frac{1}{\omega_0} \left( \mathbf{L}^T \mathbf{L} \right)^{-1} \mathbf{L}^T \left(\begin{array}{c} \dot a_1 \\ \dots \\ \dot a_n \end{array} \right), 
\end{equation}
where $\mathbf{L}$ denotes the matrix in the left-hand side of Eq.~(\ref{eqn:epsilon_system}) and $\mathbf{L}^T$ its transpose.

The determination of $\nu$ and $\epsilon$ is done following the same approach. By Eq.~(\ref{eqn:equa_omega}), and remembering that $\omega_{\infty} \simeq \omega_0 (1+ \alpha / 2)$:
\begin{equation}\label{eqn:epsilon_system2}
\omega_0 \left(\begin{array}{ccc} \frac{4a_1}{3\pi} & \frac{3a_1^2}{8} \\  \dots \\ \frac{4a_n}{3\pi} & \frac{3a_n^2}{8}  \end{array} \right)  \left(\begin{array}{c} \nu \\ \epsilon \end{array} \right) = \left(\begin{array}{c} \Delta\omega_1 \\ \dots \\ \Delta\omega_n \end{array} \right).
\end{equation}
where $\Delta\omega_i = \omega_i - \omega_{\infty}$ is the frequency shift. The solution in the least mean square sense is given by:
\begin{equation}\label{eqn:epsilon_conservative}
\left(\begin{array}{c} \nu \\ \epsilon \end{array} \right) = \frac{1}{\omega_0} \left( \mathbf{M}^T \mathbf{M} \right)^{-1} \mathbf{M}^T \left(\begin{array}{c} \Delta\omega_1 \\ \dots \\ \Delta\omega_n \end{array} \right),
\end{equation}
where $\mathbf{M}$ denotes the matrix of Eq.~(\ref{eqn:epsilon_system2}).


The five coefficients $\lambda$, $\zeta$, $\delta$, $\nu$ and $\epsilon$ are determined by Eqs.~(\ref{eqn:epsilon_dissipative}), (\ref{eqn:epsilon_conservative}). The coefficient $\alpha$ is determined as $\alpha= \omega_{\infty}^2 / \omega_{0}^2 - 1$.

Note that the above method, based on linear matrix equations, is qualitatively different from the classical decrement method. In addition, it is not limited to the six forces previously discussed. Any other nonlinear force $f=\gamma h(\theta,\dot\theta/\omega_0)$ may easily be included in the analysis by calculating its signature on the envelope and frequency shift. The terms $\omega_0\gamma \langle h\sin\varphi\rangle$ and $\omega_0\gamma \langle h\cos\varphi\rangle$ (Appendix) must respectively be added to the right-hand sides of Eqs.~(\ref{eqn:equa_diff_a}) and (\ref{eqn:equa_omega}). If the force modifies the envelope, the matrix $\mathbf L$ of Eq.~(\ref{eqn:epsilon_system}) will contain a fourth column whose entries are $\langle h\sin\varphi\rangle_i$, estimated at all time $t_i$ and $\gamma$ will appear as a supplementary unknown. If it modifies the frequency shift, then $\langle h\cos\varphi\rangle_i$ at time $t_i$ constitutes a third column of $\mathbf M$ and $\gamma$ appears as unknown. More generally, a force may have a signature on both the envelope and frequency shift. This raises the question of the unicity of the signature of a given force on envelope and frequency. In general, there is no such unicity. For instance, both forces $\dot\theta^2\sgn(\dot\theta)$ and $\theta^2\sgn(\dot\theta)$ have the exact same mean values $\langle h\sin\varphi\rangle$ and $\langle h\cos\varphi\rangle$, making them indistinguishable by the method presented here.

\section{Experimental illustration}
\label{sec:experiment}
We illustrate here the method presented above on the example of a measurement performed with the torsional magnetic levitation tribometer described in~\cite{bib:vasko2018}. The natural moment of inertia of the pendulum is $I_0=1.24$~kg$\cdot$mm$^2$. A mass $m=0.90$~g has been added at a distance $l=10.0$~mm from its rotation axis, yielding an effective moment of inertia $I=I_0+ml^2=1.33$~kg$\cdot$mm$^2$.
A spherical cap of radius $R=1.17$~mm is attached at the pendulum's tip, made of Sylgard 184 PolyDiMethylSiloxane (PDMS), prepared as in~\cite{bib:sahli2018,bib:mergel2019}. The PDMS sphere is set into contact against a glass plate under normal load $P=15.9$~mN. We consider the data such that the amplitude of the oscillation decays from about 0.3~rad to 0.02~rad, when the vertical discretization of the digital signal becomes significant. 

Figure~\ref{fig:expPDMS} shows the time-evolution of $\theta$ (solid line). The coefficients $\lambda$, $\zeta$, $\delta$, $\nu$ and $\epsilon$ have been calculated by applying Eqs.~(\ref{eqn:epsilon_dissipative}) and (\ref{eqn:epsilon_conservative}), where the values of $a_i$, $\dot a_i$, and $\omega_i$ have been assessed at $76$ instants from the full measurement of $\theta(t)$. In practice, each of 78 half-periods is first fitted using a half-sine function of amplitude $b_i$, with an extremum reached at time $t_i$. We then compute $a_i=(b_{i+1}+b_{i-1})/2$, $\dot a_i=(b_{i+1}-b_{i-1}) / (t_{i+1}-t_{i-1})$ and $\omega_i=2 \pi / (t_{i+1}-t_{i-1})$. The angular frequency $\omega_{\infty}=9.91$ rad/s is taken as the last computed value of $\omega_i$, where the influence of nonlinear forces is the least. The obtained values are $\lambda=-0.000023$, $\zeta=0.0097$, $\delta=0.037$, $\nu=-0.40$, and $\epsilon=0.30$. In Fig.~\ref{fig:expPDMS}, the dash-dotted line corresponds to the numerical solution of Eq.~(\ref{eqn:mvt}) with the three dissipative forces only, $f=\lambda \sgn(\dot\theta) + 2\zeta\dot\theta/\omega_0 + \delta \dot \theta^2 \sgn(\dot \theta)/\omega_0^2$, while the dashed line corresponds to the solution with all five forces (the effect of $\alpha$ is already accounted for in $\omega_{\infty}$), $f=\lambda \sgn(\dot\theta) + 2\zeta\dot\theta/\omega_0 + \delta \dot \theta^2 \sgn(\dot \theta)/\omega_0^2 + \nu \theta^2 \sgn(\theta) + \epsilon \theta^3$. It is clear that both curves, based on either three or five forces, well capture the envelope of the measurement. However, the curve for three forces (dash-dotted line) rapidly shows a phase shift and is even out-of-phase after few periods. The curve for five forces (dashed line) does not have this shortcoming and is in perfect agreement with the measurements over the full time window. This result highlights that the presence of nonlinear elastic forces is essential to explain the frequency shift observed on the tribometer.


\begin{figure}
\includegraphics[scale=0.45]{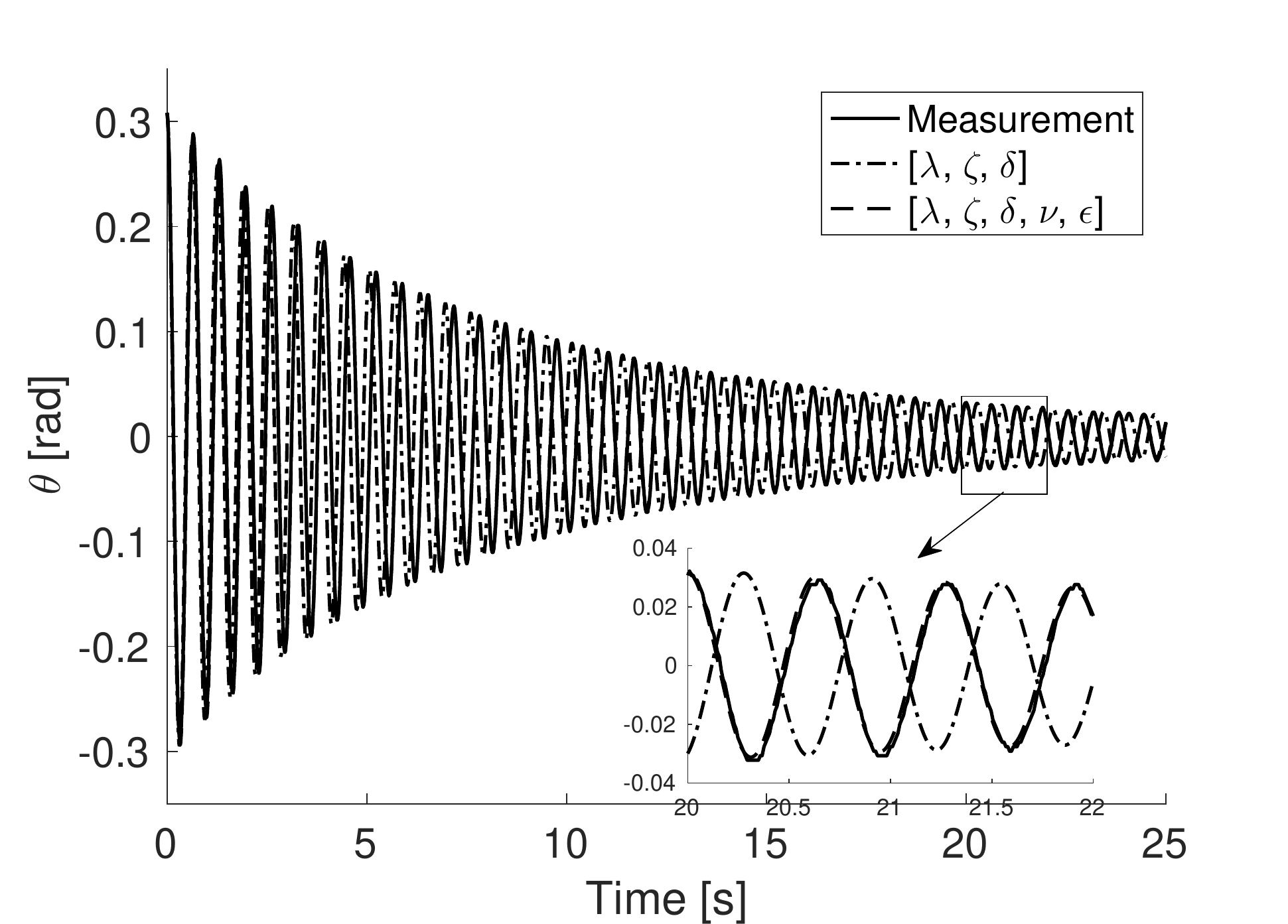}
\caption{Time evolution of $\theta$ for a PDMS-sphere/glass-plate torsional contact. Solid line: measurement. Dash-dot line: solution to Eq.~(\ref{eqn:mvt}) with $\lambda$, $\zeta$, and $\delta$. Dashed line: with $\lambda$, $\zeta$, $\delta$, $\nu$, and $\epsilon$.}
\label{fig:expPDMS}
\end{figure}


\section{Discussion}
\subsection{Values of the experimental parameters}

As shown in section~\ref{sec:strength}, $\alpha$ can be estimated from the knowledge of $\omega_0$ and $\omega_{\infty}$. With the magnetic levitation tribometer used here, accessing $\omega_0$ is not direct, because the pendulum cannot be operated in the absence of a mechanical contact. To estimate $\omega_0$, we thus performed an experiment on a contact which is expected to be submitted to a small frictional torque. We chose a contact between steel and graphite (SG), for which the measured coefficient of friction $\mu=0.05$  is the smallest among all cases investigated with this rotational tribometer. For a normal load $P=16.5$~mN, close to that used for the PDMS/glass contact, we found $\omega_{\infty,\mathrm{SG}}$=8.87~rad/s, which is an upper limit for $\omega_{0}$, if the steel/graphite contact had a vanishing frictional torque. Assuming that $\omega_0=\omega_{\infty,\mathrm{\mathrm{SG}}}$, we infer that the PDMS/glass contact torsional stiffness is estimated by $K_c\approx I(\omega_{\infty,\mathrm{PDMS}}^2 - \omega_0^2)=25.9\,10^{-6}$~N$\cdot$m. The  torsional stiffness of a non-slipping elastic sphere of radius $a$ is $K_c=16Ga^3/3$ where $G$ is the shear modulus of the elastic material~\cite{bib:johnson}. We measured $a=0.20$~mm by direct visualization through the glass and with 
$G=0.53$~MPa~\cite{bib:sahli2018}, it yields $K_c\approx22.6\,10^{-6}$~N.m, which is in reasonable agreement with the value found from the analysis of the oscillation. This agreement suggests that the change in final frequency is due to the finite torsional elasticity of the elastomer contact.

\begin{figure}[tb!]
\includegraphics[scale=0.45]{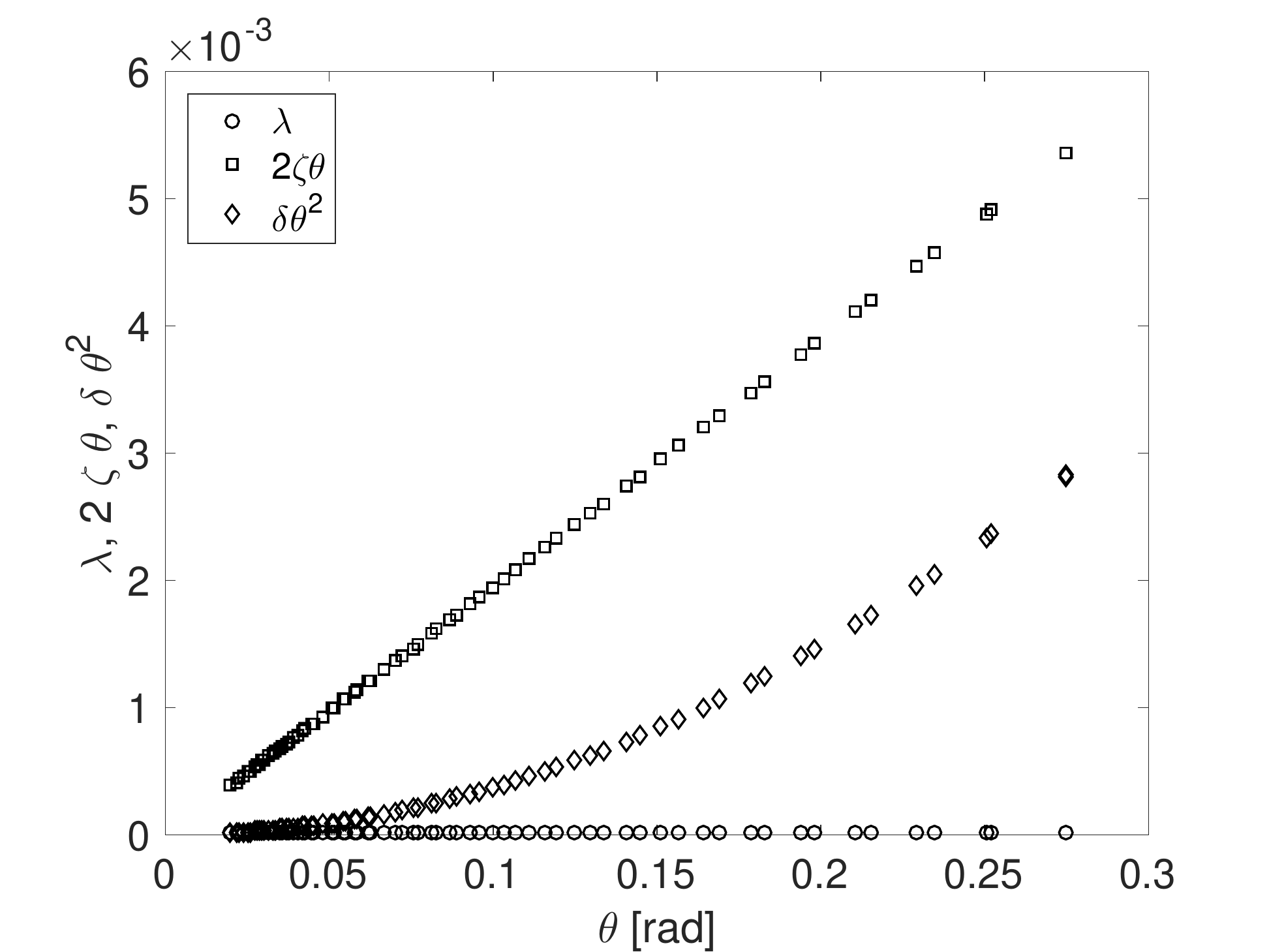}
\caption{Relative importance of solid friction force ($\lambda$), viscous force ($\zeta$), and quadratic dissipative force ($\delta$), as a function of the angle, for the experiment shown in Fig.~\ref{fig:expPDMS}.}
\label{fig:weightAmpli}
\end{figure}

Once the coefficients of all forces are known, it is interesting to assess the relative weights of those forces in the signal. Those weights depend not only on the non-dimensional coefficients, but also on the current amplitude of $\theta$ or $\dot \theta$. Approximating $\omega \simeq \omega_0$, one can estimate the weight of each of the five considered forces by: $\lambda$, $2\zeta \theta$, $\delta \theta^2$, $4\nu \theta / (3 \pi)$ and $3 \epsilon \theta^2 / 8$. As an example, in Fig.~\ref{fig:weightAmpli}, the weights of the three forces affecting the amplitude in the PDMS/glass experiment of Fig.~\ref{fig:expPDMS} are shown. For the experimental values of the coefficients and the experimental range of angles, one can see that the viscous force dominates and is thus responsible of most of the dissipation in the pendulum. \rev{Such a dominance does not necessarily indicate that the interfacial friction is strongly viscous (and indeed, \cite{bib:scheibert2009,bib:yashima2015} found negligible velocity-dependence of friction at similar PDMS/glass interfaces). Actually, this viscous force term presumably also} combines viscous dissipation in the air around the pendulum, in the viscoelastic \rev{bulk of the PDMS}, and in the magnetic device. \rev{In comparison,} the contribution of a solid-friction-like term ($\lambda$-term) is found completely negligible. The obtained value is even negative, which likely suggests that the uncertainty on $\lambda$ is larger than its value.

Figure~\ref{fig:weightPhase} shows the relative importance of the two conservative forces, $\nu\theta^2\sgn(\theta)$ and $\epsilon\theta^3$, in the same experiment. The quadratic elastic force $\nu\theta^2\sgn(\theta)$ clearly dominates and imposes the phase shift in the pendulum. This nonlinear force may be due to a combination of a nonlinear stiffness of the torsional contact, a nonlinear elastic behaviour law of the PDMS and a nonlinear restoring force due to the magnetic device.

It is interesting to compare the value of $\epsilon$ found with that expected if the pendulum was a pure pendulum under gravity. In that case, the second order approximation of $\sin \theta$, which enters the exact equation of motion of the pendulum, is $\sin\theta\simeq\theta- \theta^3 / 6$. This means that the expected value of $\epsilon$ would be $-1 / 6$. The fact that we find a value about twice bigger in magnitude and of opposite sign indicates that apart from gravity, there is a stronger cubic elastic force in the system, presumably due to the magnetic levitation device.

\begin{figure}[tb!]
\includegraphics[scale=0.45]{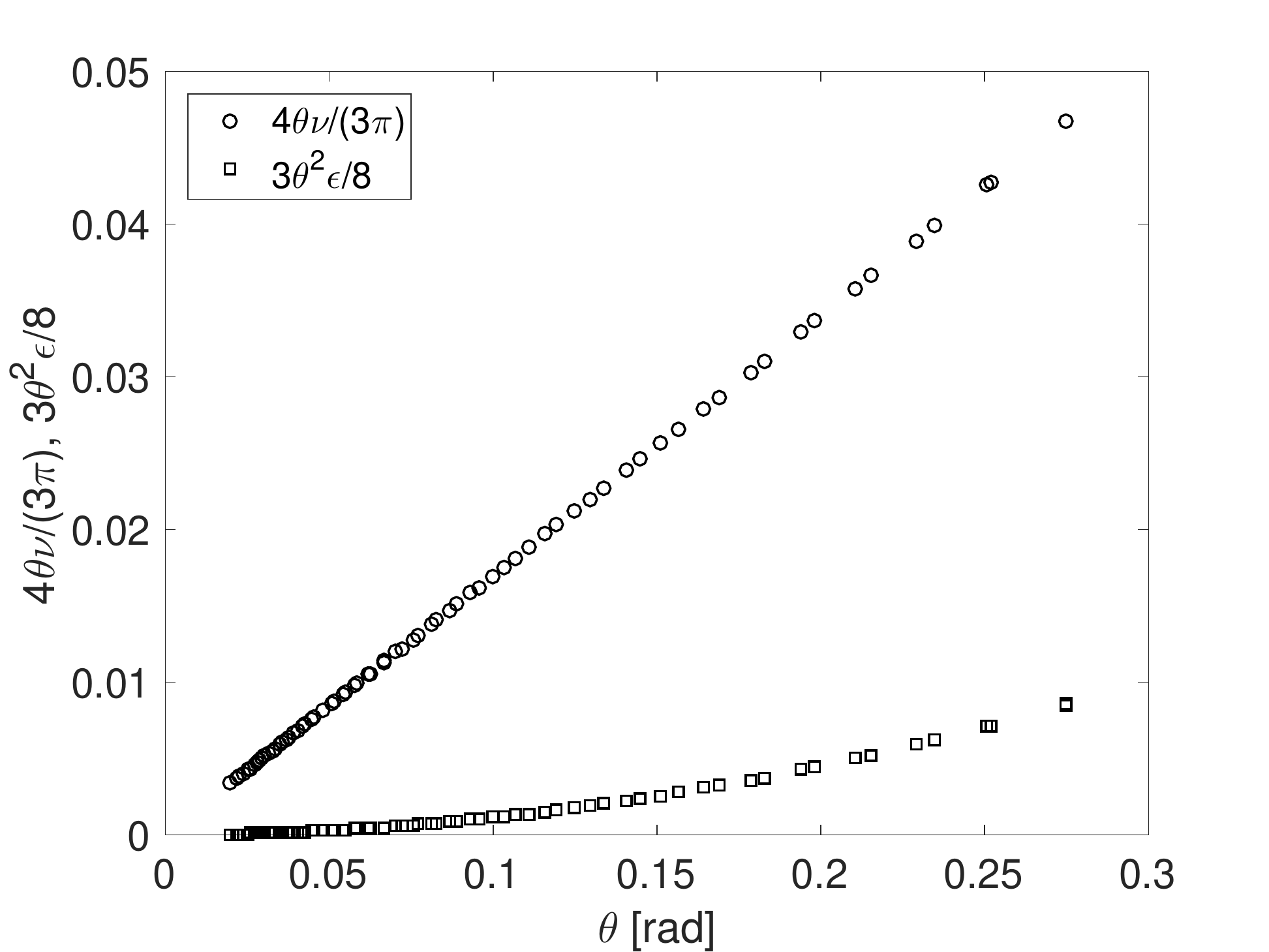}
\caption{Relative importance of quadratic ($\nu$) and cubic ($\epsilon$) elastic forces on the phase shift, as a function of the angle, for the experiment shown in Fig.~\ref{fig:expPDMS}.}
\label{fig:weightPhase}
\end{figure}

Also of interest is the instantaneous frequency $\omega$ versus the magnitude $a$ of oscillation (Fig.~\ref{fig:phaseAmpli}), for the measured curve $\theta(t)$ of Fig.~\ref{fig:expPDMS}. $\omega(a)$ is almost linear, as shown in Fig.~\ref{fig:periodQuadraticConservative}. This observation is fully consistent with the dominance of the $\nu$ term compared to the $\epsilon$ term, already demonstrated in Fig.~\ref{fig:weightPhase}. Furthermore, the negative slope of $\omega(a)$ proves that $\nu<0$. Note that a similar, affine-like behaviour of $\omega(a)$ had already been obtained, for the same tribometer, on a steel/glass contact~\cite{bib:vasko2018}, suggesting that a similar quadratic elastic force was also important in that case.

\begin{figure}[tb!]
\includegraphics[scale=0.45]{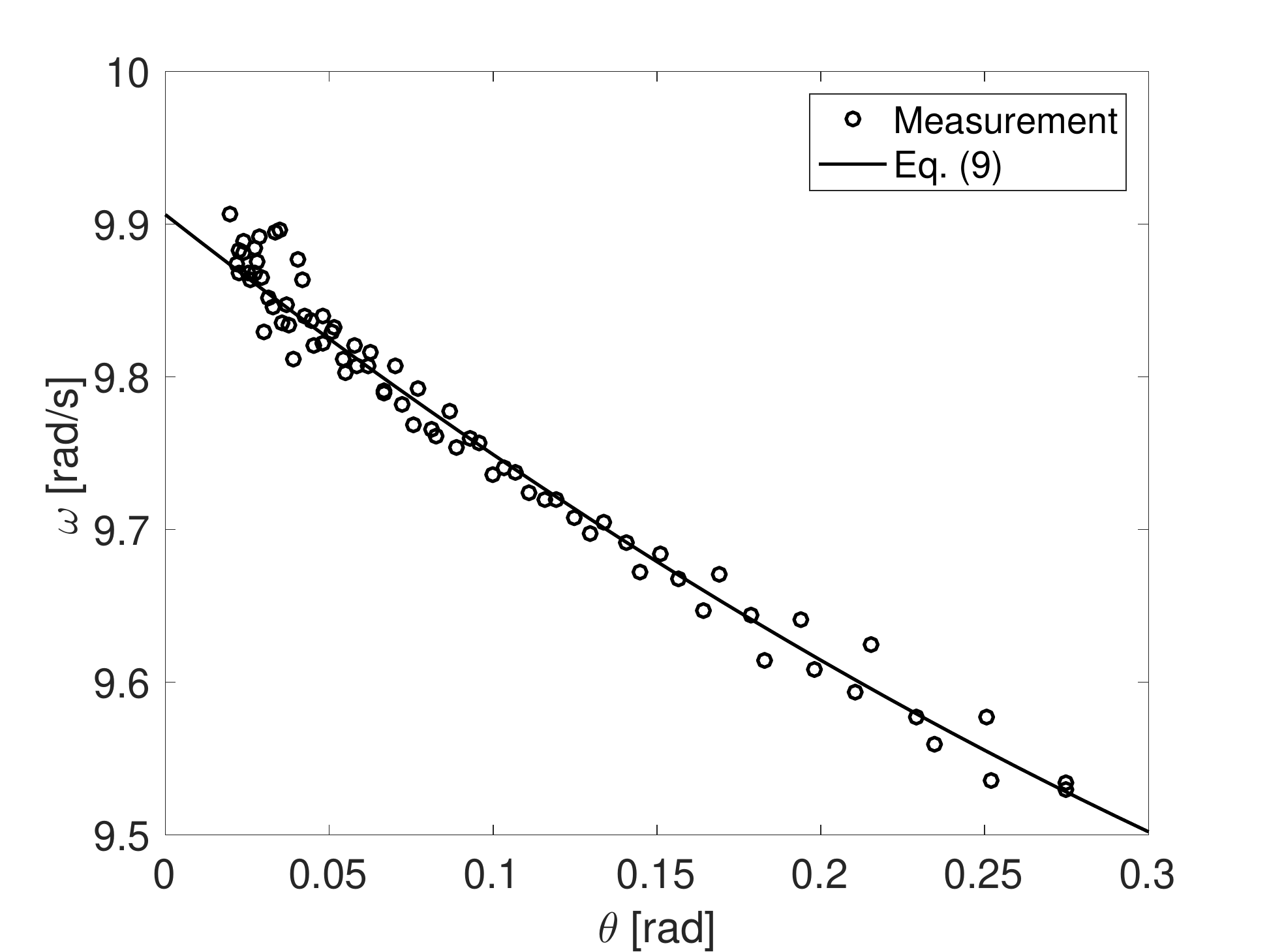}
\caption{Evolution of angular frequency, $\omega$, versus magnitude of oscillation, $a$, for the experiment shown in Fig.~\ref{fig:expPDMS}.}
\label{fig:phaseAmpli}
\end{figure}

\subsection{General comments}


The analysis in terms of the weights of the various forces reveals several generic features. Concerning the decay of amplitude, the quadratic term $\delta \theta^2$ will always dominate at large amplitude, at the beginning of an experiment. In particular, it will be larger than the viscous term for amplitudes $a > 2 \zeta / \delta$ (except of course if this value exceeds $\pi$, the maximum possible initial angle). For our experiment of Figs.~\ref{fig:expPDMS} and~\ref{fig:weightAmpli}, it would correspond to amplitudes above about 0.5.

Another general results is that, at small amplitudes, the constant $\lambda$-term will always dominate. This means that in any situation in which $\lambda\neq0$, the oscillation will vanish at a finite time. A rough estimate of this arrest time can be obtained as follows. At small amplitudes, the $\lambda$-term will dominate when the amplitude $a$ will become smaller that $a_{\lambda}=\lambda / (2 \zeta)$ (equating $\lambda$ and $2 \zeta a$). After that instant, assuming that $\lambda$ is the only term important for the amplitude decay, one can use Eq.~(\ref{eqn:env_friction}) to estimate the remaining oscillation time to be $\pi a_{\lambda} / (2 \lambda \omega_0)= \pi / (4 \zeta \omega_0)$. So, after the amplitude $a_{\lambda}$ is reached, the oscillation will cease after a time of the order of $\pi / (4 \zeta \omega_0)$.

Similarly, concerning the frequency shift, the $\epsilon$-term, which is quadratic in amplitude, will always dominate the $\nu$-term for amplitudes $a>32 \nu / (9 \pi \epsilon)$. For our experiment of Figs.~\ref{fig:expPDMS} and~\ref{fig:weightPhase}, it would correspond to amplitudes above about 4.76. This value is larger than $\pi$, the largest possible initial angle, which means that, in our experimental case, event for the largest amplitudes, $\omega(a)-\omega_0$ will always remain dominantly linear. In general, the linear $\nu$-term will always dominate at small amplitudes, when the angular measurement may become less accurate, due to discretizaton effects. Our conclusion is that, in this regime, a linear extrapolation of $\omega(a)$ for small $a$ is \textit{a priori} a correct approximation of the behaviour of the system.

\rev{More generally, friction being an hysteretic phenome\-non, it cannot always be described only as a function of the instantaneous angle and angular velocity, as assumed in Eq.~(\ref{eqn:mvt}). In transient regimes, like the oscillations considered in relaxation tribometry, a more complete description of the interface should incorporate one or several state variables, as in the rate-and-state (RS) friction framework~\cite{bib:ruina1983,bib:baumberger2006}. Although most RS models use a typical contact-related time as a state variable~\cite{bib:barsinai2012,bib:tromborg2014,bib:thogersen2014,bib:hatano2015}, here we believe that the most relevant state variable would be related to the oscillating history of the contact. When the contact is brought to its initial angle, only a central circular part of radius $c_0$ of the contact has remained in a stuck state, while its periphery has already been slipping~\cite{bib:chateauminois2010}. When the contact has completed its first half-cycle of oscillation and is back to vanishing angular velocity, it has  an angle the absolute value of which is smaller than the initial one, and thus the stick radius is now $c_1>c_0$. As a consequence, when the oscillator goes back to a vanishing angle, the annulus between $c_1$ and $c_0$ in the contact region has stored a shear strain state which is different from that of the first contact, and which will survive all along the subsequent decaying oscillation of the contact. Such a scenario occurs over and over at each half-cycle, building up a complex, onion-ring-like shear strain field, the description of which involves knowledge of the series of $c_i$ reached at all half-cycles. Incorporating such a complex state variable in the analysis of relaxation tribometry data is an interesting future challenge. It will likely require extension of studies limited to the first loading of elastic contacts~\cite{bib:johnson,bib:chateauminois2010,bib:prevost2013}, to decaying oscillations.}

\section{Conclusion}
\label{sec:conclusion}
We have shown that relaxation tribometry is not limited to the measurement of constant friction and viscous coefficients as is usually done using the decrement method. More complex dissipative forces but also elastic forces can be unambiguously identified and quantified using the general procedure proposed in this study. The key is to exploit the two-times averaging method to analyse not only the time-evolution of the vibration decay, but also that of the frequency shift. The magnitudes of forces are then solutions of a linear system, although the forces are themselves nonlinear. This procedure, which has been applied to six relevant types of contact forces, can easily be extended to any other desired nonlinear force, to identify its characteristic signature, both on the amplitude and frequency. Those results suggest that relaxation tribology has a vast, but still insufficiently exploited potential, both fundamental (identification of the forces at play) and applied (quantification of those forces).

\appendix
\section{two-times averaging method}\label{sec:multiscale}
The right-hand side of Eq.~(\ref{eqn:mvt}) is written $f=\epsilon h(\theta,\theta')$ where $\epsilon<<1$ and $\theta'=\dot\theta/\omega_0$ denotes the derivative of $\theta$ with respect to dimensionless time. We seek the solution of the form $\theta=a \cos \varphi$, where $\varphi = \omega_0 t + \phi$ and $a(t)$ and $\phi(t)$ are slowly varying functions. Substituting in Eq.~(\ref{eqn:mvt}) gives:
\begin{equation}
\begin{array}{l}
\ddot a \cos\varphi - 2\dot a (\omega_0 + \dot\phi)\sin\varphi - a\ddot \phi \sin\varphi - 2a\dot\phi \omega_0\cos\varphi \\
 - a \dot\phi^2\cos\varphi = -\epsilon \omega_0^2 h.\label{Eq:App1}
 \end{array}
\end{equation}
Considering that $\ddot a$, $\dot a \dot\phi$, $\ddot\phi$, and $\dot\phi^2$ are second order terms in $\epsilon$ and can thus be neglected in Eq.~(\ref{Eq:App1}):
\begin{equation}\label{eqn:App2}
 2\dot a \omega_0 \sin\varphi + 2a \dot \phi \omega_0 \cos\varphi = \epsilon \omega_0^2 h.
\end{equation}
We must now develop $h$ at order zero in $\epsilon$ since the left-hand side if of order one in $\epsilon$. At order $0$, $\theta = a\cos\varphi$ and $\dot \theta/\omega_0 = -a\sin\varphi$ therefore $h=h(a\cos\varphi,-a\sin\varphi)$.
Then substituting in Eq.~(\ref{eqn:App2}) and averaging over a time-period (with $\dot a$ and $\dot \phi$ constant) gives the so-called averaged equations:
\begin{eqnarray}\label{eqn:hsin}
\dot a & = & \frac{\epsilon\omega_0}{2\pi} \int_0^{2\pi} \!\!\!\!\!\!\! h(a\cos\varphi,-a\sin\varphi)\sin\varphi\,\dd\varphi = \omega_0 \epsilon \langle h(\varphi) \sin\varphi \rangle \\
\label{eqn:hcos}
a\dot \phi & = & \frac{\epsilon\omega_0}{2\pi} \int_0^{2\pi} \!\!\!\!\!\!\! h(a\cos\varphi,-a\sin\varphi)\cos\varphi\,\dd\varphi = \omega_0 \epsilon \langle h(\varphi) \cos\varphi \rangle
\end{eqnarray}
where $\langle \cdot \rangle$ denotes mean value over $2\pi$. These are two first-order ordinary differential equations on $a$ and $\phi$.

For instance, consider the case of a quadratic dissipative force $f=\epsilon\dot\theta^2\sgn(\dot\theta)/\omega_0^2$. Then $h(\theta,\theta')= \theta'^2\sgn(\theta')$ and $h(\varphi) = -a^2 \cos^2 \varphi \sgn(\sin\varphi)$. By averaging:
\begin{eqnarray}\label{eqn:App3}
\langle h \sin\varphi \rangle & = & -\frac{4a^2}{3\pi}, \\
\langle h \cos\varphi \rangle & = & 0.
\end{eqnarray}
 The two differential equations are therefore $\dot a = -4a^2 \epsilon\omega_0 / 3\pi$ and $\dot \phi = 0$. After integration, $a(t) = \left[a_0^{-1}+4\epsilon \omega_0 t / 3\pi\right]^{-1}$ (Eq.~\ref{eqn:env_quadDiss}) and $\phi(t) = \phi_0$, $a_0$ and $\phi_0$ being the initial values of $a$ and $\phi$.

A second interesting example is $f=\epsilon\theta^2\sgn(\dot\theta)$ for which $h(\theta,\theta')=\theta^2\sgn(\theta')$ and $h(\varphi) = -a^2 \cos^2 \varphi \sgn(\sin\varphi)$. By averaging:
\begin{eqnarray}
\langle h \sin\varphi \rangle & = & -\frac{2a^2}{3\pi}, \\
\langle h \cos\varphi \rangle & = & 0,
\end{eqnarray}
which gives the same signature $\dot a \propto a^2$ as in Eq.~(\ref{eqn:App3}).

Similar results for all considered contact forces are summarized in Table~\ref{tab:1}.

\begin{table}
\caption{Averaged values of $h$ of Eqs.~(\ref{eqn:hsin}) and (\ref{eqn:hcos}) for the six considered forces}
\label{tab:1}       
\begin{tabular}{lllllll}
\hline\noalign{\smallskip}
$h$ & $\sgn(\theta')$ & $\theta'$ & $\theta'^2\sgn\theta'$ & $\theta$ & $\theta^2\sgn\theta$ & $\theta^3$\\
\noalign{\smallskip}\hline\noalign{\smallskip}
$\langle h\sin\varphi \rangle$ & $-\frac{2}{\pi}$ & $-\frac{a}{2}$ & $-\frac{4a^2}{3\pi}$ & $0$ & $0$ & $0$ \\
$\langle h\cos\varphi \rangle$ & $0$ & $0$ &  $0$ & $\frac{a}{2}$ & $\frac{4a^2}{3\pi}$ & $\frac{3a^3}{8}$ \\
\noalign{\smallskip}\hline
\end{tabular}
\end{table}

\begin{acknowledgements}
We thank J. Perret-Liaudet, E. Rigaud, and O.A. Marchenko for fruitful discussions and critical comments. This work was supported by CNRS-Ukraine PICS Grant No. 7422. 
\end{acknowledgements}



\end{document}